\documentclass[epj]{svjour}
\usepackage{graphics}
\begin{document}
\title{Non-trivial effect of the in-plane shear elasticity on the phase transitions of fixed-connectivity meshwork models}


\author{Isao Endo\dag\ and Hiroshi Koibuchi\ddag\ 
}

\institute{\dag\ Department of Electrical and Systems Engineering, \\
 Ibaraki National College of Technology, 
 Nakane 866, Hitachinaka,  Ibaraki 312-8508, Japan\\
\ddag\ Department of Mechanical and Systems Engineering \\
  Ibaraki National College of Technology \\
  Nakane 866, Hitachinaka, Ibaraki 312-8508, Japan}

%
%
\abstract{
We numerically study the phase structure of two types of triangulated spherical surface models, which includes an in-plane shear energy in the Hamiltonian, and we found that the phase structure of the models is considerably influenced by the presence of the in-plane shear elasticity. The models undergo a first-order collapsing transition and a first-order (or second-order) transition of surface fluctuations; the latter transition was reported to be of second-order in the first model without the in-plane shear energy. This leads us to conclude that the in-plane elasticity strengthens the transition of surface fluctuations. We also found that the in-plane elasticity decreases the variety of phases in the second model without the in-plane energy. The Hamiltonian of the first model is given by a linear combination of the Gaussian bond potential, a one-dimensional bending energy, and the in-plane shear energy. The second model is obtained from the first model by replacing the Gaussian bond potential with the Nambu-Goto potential, which is defined by the summation over the area of triangles. 
}
\authorrunning {I.Endo and H.Koibuchi}
\titlerunning {Non-trivial effect of the in-plane shear elasticity on the phase transitions}

\maketitle
\section{Introduction}
Surface models are conventionally defined by a surface tension energy and a bending energy \cite{HELFRICH-1973,POLYAKOV-NPB1986,KLEINERT-PLB1986}. Both of the energies play a role of maintaining the shape of surface against environmental external forces including thermal fluctuations \cite{NELSON-SMMS2004-1,David-TDQGRS-1989,Wiese-PTCP2000,Bowick-PREP2001,Gompper-Schick-PTC-1994,WHEATER-JP1994}, Thus, the surface models are always constructed to have resistance against tensile deformations and bending deformations, while no resistance is assumed against in-plane shear deformations. This final assumption seems valid in fluid membranes at least, because the in-plane shear deformation has no cost in energy in the fluid membranes.

It was reported that a collapsing transition and a transition of surface fluctuations occur in the conventional fixed connectivity surface models, where no in-plane shear energy is included in the Hamiltonians \cite{KOIB-PRE-20045-NPB-2006}. These two transitions are identical to the so-called crumpling transition, which has been extensively studied so far \cite{KANTOR-NELSON-PRA1987,AMBJORN-NPB1993,Peliti-Leibler-PRL1985,DavidGuitter-EPL1988,PKN-PRL1988}.

However, it is unclear whether the in-plane shear elasticity is negligible to the bending elasticity in membranes in the gel phase or in membranes supported by cytoskeletons. In fact, red blood cells are known to have non-negligible in-plane shear elasticity \cite{HLRG-BIOPJ-1999,LHRSG-BIOPJ-2001}.

Therefore, it is natural to ask whether the in-plane shear elasticity influences the transitions, which can be seen in the models defined without the in-plane shear energy. In this article, we study two types of meshwork models in \cite{KOIB-PLA2007} and \cite{KOIB-EPJB2007-3} by including an in-plane shear energy in the Hamiltonian in order to see possible influences of the in-plane elasticity on the phase transitions.

The first meshwork model, studied in \cite{KOIB-PLA2007}, is distinguished from the conventional surface models because of the difference in the phase structures of them. The meshwork model undergoes a first-order collapsing transition between the collapsed phase and the smooth phase and a continuous transition of surface fluctuations at the same transition point, while both of the transitions are of first-order in the conventional models. The transition of surface fluctuations distinguishes the meshwork model in \cite{KOIB-PLA2007} from the conventional surface models such as those in \cite{KOIB-PRE-20045-NPB-2006}.

The second meshwork model, studied in \cite{KOIB-EPJB2007-3}, is obtained from the model in \cite{KOIB-PLA2007} by replacing the Gaussian bond potential with the Nambu-Goto potential, which is given by the summation over the area of triangles. The Nambu-Goto surface model is well-known as an ill-defined model in the sense that the model has no smooth surface in the whole range of the bending energy $b$ when the Hamiltonian includes the conventional two-dimensional bending energy of the type $1\!-\!{\bf n}\cdot {\bf n}$ as the curvature energy, where ${\bf n}$ is a unit normal vector of a triangle \cite{ADF-NPB1985}. The well-definedness of the model in \cite{KOIB-EPJB2007-3} is due to the one-dimensional bending energy, and a variety of shapes makes the model very different from the conventional ones in \cite{KOIB-PRE-20045-NPB-2006}, which have only the smooth phase and the collapsed phase. The models in \cite{KOIB-PLA2007,KOIB-EPJB2007-3} are also different form the surface models with cytoskeletal structures in \cite{KOIB-PRE2007,KOIB-JSTP-2007-1}, because the length $L$ between the junctions of the models in \cite{KOIB-PLA2007,KOIB-EPJB2007-3} is $L\!=\!1$ in the unit of bond length while that of the models in  \cite{KOIB-PRE2007,KOIB-JSTP-2007-1} is $L\!>\!1$.

We should note that the linear bending energy assumed in the meshwork models in \cite{KOIB-PLA2007} and \cite{KOIB-EPJB2007-3} produces no resistance force against the in-plane deformations of the surface; the junctions of the meshwork play only a role for binding the one-dimensional skeletons. This makes us to expect that the in-plane energy, as an additional energy term in the Hamiltonian, has a non-trivial influence on the phase transitions of the models. For this reason, we study in this paper the meshwork models in \cite{KOIB-PLA2007} and \cite{KOIB-EPJB2007-3} by including the in-plane energy in the Hamiltonians. 

Our main conclusion in this paper is that the in-plane shear energy enhances the surface fluctuations. We should comment on why the in-plane energy makes the transitions strong. The reason seems that the energy of thermal fluctuations is accumulated mainly in the bending deformation energy of the surface if the surface has large in-plane shear resistance. Since the surface deformation can be divided into the bending deformation and the in-plane deformation, then the bending deformation is enhanced if the in-plane deformation is suppressed. 

\section{Models and Monte Carlo technique}
The triangulated lattices, on which two types of models are defined, are identical to those in \cite{KOIB-PLA2007,KOIB-EPJB2007-3}. By splitting the icosahedron, we have a triangulated spherical meshwork of size $N\!=\!10\ell^2\!+\!2$, which is the total number of vertices. The symbol $\ell$ in $N$ is the number of partitions of an edge of the icosahedron. The co-ordination number $q$, which is the total number of bonds emanating from a vertex, is $q\!=\!5$ on $12$ vertices, which correspond to those of the icosahedron, and $q\!=\!6$ on the remaining $N\!-\!12$ vertices. 

The Hamiltonian $S$ is given by a linear combination of the bond potential $S_1$, the one-dimensional bending energy $S_2$, and the in-plane shear energy $S_3$, which are defined by
\begin{eqnarray}
\label{Disc-Eneg} 
&&S_1= \left\{
       \begin{array}{@{\,}ll}
       S_{1,\rm G}  & \quad ({\rm model \; 1} ), \\ \nonumber
       S_{1,\rm NG} & \quad ({\rm model \; 2}), 
       \end{array}
       \right.
       \quad S_2=\sum^\prime_{(ij)}\left(1-{\bf t}_i\cdot {\bf t}_j\right), \nonumber \\ 
&&S_3=\sum_i \left[1-\cos(\delta_i-\pi/3)\right].
\end{eqnarray} 
$S_{1,\rm G}$ and $S_{1,\rm NG}$ denote the Gaussian bond potential and the Nambu-Goto potential, which are respectively given by 
\begin{equation}
\label{Bond-potential} 
S_{1,\rm G}=\sum_{(ij)} \left(X_i-X_j\right)^2,\quad  
S_{1,\rm NG}=\sum_{\mit \Delta} A_{\mit \Delta}.
\end{equation} 
The symbol $X_i$ in $S_{1,\rm G}$ denotes the three-dimensional position of the vertex $i$, and $ \sum_{(ij)} $ in $S_{1,\rm G}$ denotes the sum over bonds $(ij)$, which connect the vertices $i$ and $j$. $A_{\mit \Delta}$ in $S_{1,\rm NG}$ denotes the area of the triangle ${\mit \Delta}$. We call the model defined by $S_1\!=\!S_{1,\rm G}$, $S_2$, and $S_3$ as {\it model 1}, and the one defined by $S_1\!=\!S_{1,\rm NG}$, $S_2$, and $S_3$ as {\it model 2}.

\begin{figure}[hbt]
\centering
\resizebox{0.46\textwidth}{!}{%
\includegraphics{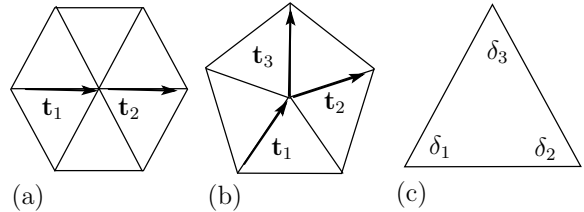}
}
\caption{(a) Unit tangential vectors ${\bf t}_1$ and ${\bf t}_2$ at a vertex of coordination number $q\!=\!6$, and (b) those at a vertex of coordination number $q\!=\!5$, and (c) three internal angles  $\delta_i(i\!=\!1,2,3)$ in the definition of $S_3$. The vectors in (a) and those in (b) define the bending energy $1\!-\!{\bf t}_1\cdot {\bf t}_2$ and $[1\!-\!{\bf t}_1\cdot ({\bf t}_2\!+\!{\bf t}_3)/2]/2$, respectively.}
\label{fig-1}
\end{figure}
In $S_2$ of Eq. (\ref{Disc-Eneg}), ${\bf t}_i$ is a unit tangential vector of the bond $i$. The definition of $\sum^\prime_{(ij)}$ in $S_2$ is identical to that in \cite{KOIB-PLA2007,KOIB-EPJB2007-3} and is summarized as follows:  $1\!-\!{\bf t}_1\cdot {\bf t}_2$ is included in $S_2$ with the weight of $1$ at the vertex of $q\!=\!6$ shown in Fig. \ref{fig-1}(a), while $1\!-\!{\bf t}_1\cdot (\!{\bf t}_2 \!+\!{\bf t}_3)/2$ is included in $S_2$ with the weight of $1/2$ at the vertex of $q\!=\!5$ shown in Fig. \ref{fig-1}(b). As a result, we have $\sum^\prime_{(ij)}1\!=\!N_B$, where $N_B\!=\!3N-6$ is the total number of bonds.

$\delta_i$ in $S_3$ of Eq. (\ref{Disc-Eneg}) denotes one of the three internal angles of a triangle such as shown in Fig. \ref{fig-1}(c). Thus, the summation $\sum_i$ in $S_3$ gives $\sum_i1\!=\!3N_T$, where $N_T(\!=\!2N\!-\!4)$ is the total number of triangles. Although the angles $\delta_i(i\!=\!1,2,3)$ are linearly dependent to each other, we use three of them in $S_3$ because of its non-linear definition with respect to the angle $\delta$. 

The models; model 1 and model 2, are defined by the following partition function:
\begin{eqnarray} 
\label{Part-Func}
&&Z = \int^\prime \prod _{i=1}^{N} d X_i \exp\left[-S(X)\right], \\
&&S(X)=S_1 + b S_2 + \alpha S_3,  \nonumber \\
&&S_1=S_{1,\rm G}\,({\rm model \; 1}), \;\; {\rm or}\;\; S_1=S_{1,\rm NG}\,({\rm model \; 2}),\nonumber
\end{eqnarray}
where $b$ is the bending rigidity, and $\alpha$ is the parameter denoting the in-plane rigidity of the surface. In the limit of $\alpha\to\infty$, we expect that the surface is composed of only regular triangles. On the contrary, the meshwork models in \cite{KOIB-PLA2007} and \cite{KOIB-EPJB2007-3} are restored in the limit of $\alpha\to 0$. The prime in $\int ^\prime$ denotes that the integration is performed under the condition that the center of mass of the surface is fixed.

We note that both $b$ and $\alpha$ in Eq. (\ref{Part-Func}) are the microscopic quantities and not always identical to the macroscopic ones. So it is unclear to what value $\alpha$ should be fixed in the simulations to see influence of the in-plane elasticity on the transitions. However, as mentioned in the introduction, we know that there is a membrane whose in-plane shear modulus is not negligible. We should remind ourselves of that the macroscopic bending modulus and the macroscopic shear modulus are of the same order in the red cells \cite{HLRG-BIOPJ-1999,LHRSG-BIOPJ-2001}. Therefore, we assume the value of $\alpha$ in the simulations as $\alpha\!=\!1$, which is of the same order as that of $b\!=\!b_c$ the collapsing transition point. 

The surface has no in-plane shear resistance against the in-plane shear deformations in the case of $\alpha\!\to\!0$ in both models. The variation of the internal angles $\delta$ in Fig. \ref{fig-1} changes neither $S_1$ nor $S_2$. It is easy to understand that only $S_3$ can reflect the variation of $\delta$. In fact, thin and oblong triangles form the surfaces in the planar phase and in the linear phase in the model of \cite{KOIB-EPJB2007-3}. The vertices of the surface in the models of \cite{KOIB-PLA2007,KOIB-EPJB2007-3} have only a role for binding the one-dimensional skeletons. To the contrary, the surface has the in-plane shear resistance in the case that the junctions are elastic or rigid \cite{KOIB-JSTP-2007-1}.  

The dynamical variables $X$ in $Z$ are integrated over by the canonical Metropolis Monte Carlo technique. A random shift $X\to X^\prime \!=\!X\!+\!\delta X$ is accepted with the probability ${\rm Min}[1,\exp(-\delta S)]$, where $\delta S\!=\!S({\rm new})\!-\!S({\rm old})$. The vector $\delta X$ is chosen randomly in a sphere, whose radius is fixed at the beginning of the simulations for maintaining about $50\%$ acceptance rate. 

Total number of Monte Carlo sweeps (MCS) after the thermalization MCS at the transition region of model 1 is about $(2\!\sim\! 2.5)\times 10^8$ for the $N\!=\!2562$ surface, $(4\!\sim\! 5)\times 10^8$ for the $N\!=\!4842$ surface, $(6\!\sim\! 7)\times 10^8$ for the $N\!=\!10242$ surface, and  $(8\!\sim\! 9)\!\times\! 10^8$ for the $N\!=\!16812$ surface. Relatively smaller number of MCS is performed at non-transition regions of $b$. The total number of MCS for model 2 is almost identical to or slightly smaller than that of model 1.

The standard error $\delta _Q$ of the quantity $Q$ is defined by the so-called the binning analysis: Let $N_{\rm tot}$ be the total number of data in the sequence $\{Q\}_i (i\!=\!1,...,N_{\rm tot})$, where each $Q_i$ is obtained at every $1000$ MCS. The total number of MCS performed is thus given by $1000 N_{\rm tot}$. The series $\{Q\}_i$ is split into $n_b$ sub-series $\{Q\}_I (I\!=\!1,...,n_b)$, and $Q_I$ denotes the mean value of the $I$-th sub series. The total number of MCS $n_{\rm tot }$ in a sub-series is given by $n_{\rm tot }\!=\!1000 N_{\rm tot}/n_b$. Let $Q$ denotes the mean value of $\{Q\}_I$; $Q$ is also the mean value of $\{Q\}_i$. Then, $\delta_Q$ of $\{Q\}_I$ is defined by $\delta_Q\!=\!\sqrt{\sum_{I=1}^{n_b}(Q_I\!-\!Q)^2/n_b}$, and $\delta_Q$ is expected to decrease with increasing $n_b$ if $n_{\rm tot}$ is sufficiently large such that $\{Q\}_I$ are statistically independent. The total number of MCS $n_{\rm tot }$ in a series after the thermalization MCS is assumed as follows: $n_{\rm tot }\!=\!1\!\times\!10^7$ for the $N\!=\!2562$ surface and  $n_{\rm tot }\!=\!2\!\times\!10^7$ for the $N\!=\!4842$, $N\!=\!10242$, and $N\!=\!16812$ surfaces. The assumed value $n_{\rm tot }$ for the large surfaces is insufficient and then,  $\delta _Q$ remains large.  

\section{Results}
\subsection{Gaussian bond potential model}
\begin{figure}[hbt]
\centering
\resizebox{0.49\textwidth}{!}{%
\includegraphics{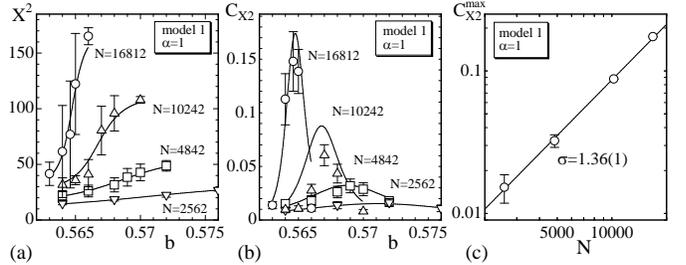}
}
\caption{(a) The mean square size $X^2$ versus $b$ of model 1, (b) the variance $C_{X^2}$ of $X^2$ versus $b$, and (c) log-log plots of the peak $C_{X^2}^{\rm max}$ against $N$. The solid lines in (a) and (b) are drawn by the multihistogram reweighting technique \cite{Janke-histogram-2002}. The straight line in (c) is obtained by fitting the data to Eq. (\ref{scaling-CX2}). The critical exponent $\sigma\!=\!1.36(1)$, which is given by the slope of the fitted line, indicates that the collapsing transition is of first-order. }
\label{fig-2}
\end{figure}

In this subsection, the numerical data of model 1 will be presented. First we show in Fig. \ref{fig-2}(a) the mean square size $X^2$ defined by 
\begin{equation}
\label{X2}
X^2={1\over N} \sum_i \left(X_i-\bar X\right)^2, \quad \bar X={1\over N} \sum_i X_i,
\end{equation}
where $\bar X$ is the center of mass of the surface. The solid lines are drawn by the multihistogram reweighting technique \cite{Janke-histogram-2002}. Rapid change of $X^2$ against $b$ can be seen in the figure as $N$ increases, and this indicates the existence of the collapsing transition between the smooth phase and the collapsed phase. The transition appears to be discontinuous, because $X^2$ seems to change discontinuously on the largest surface.

The variance of $X^2$ is defined by
\begin{equation}
\label{fluctuation-X2}
C_{X^2} = {1\over N} \langle \; \left( X^2 \!-\! \langle X^2 \rangle\right)^2\rangle,
\end{equation}
and this reflects how large the fluctuation of $X^2$ is. We show $C_{X^2}$ versus $b$ in Fig. \ref{fig-2}(b) in order to see the order of the transition more clearly. The solid lines are drawn also by the multihistogram reweighting technique. Sharp peaks seen on $C_{X^2}$ imply a phase transition. Therefore, we plot the peak values $C_{X^2}^{\rm max}$, which are obtained by the multihistogram reweighting technique, in a log-log scale against $N$ in Fig. \ref{fig-2}(c). The error bars in Figs. \ref{fig-2}(a) denote the standard errors $\delta_{X^2}$,  while the errors  in Fig. \ref{fig-2}(b) are given by $\delta_{C_{X^2}}/\sqrt{n_b}$; $\delta_{C_{X^2}}$ is too large to show, hence we divide it by $\sqrt{n_b}$.  The error bars in Fig. \ref{fig-2}(c) denote the standard errors $\delta_{C_{X^2}}$ obtained by the multihistohgram technique. The fitting of data $C_{X^2}$ are performed by using those errors in Fig. \ref{fig-2}(c).

The straight line in Fig. \ref{fig-2}(c) is drawn by the power law fitting of the data, and we have a critical exponent $\sigma$ such that 
\begin{equation}
\label{scaling-CX2}
C_{X^2}^{\rm max} \propto N^\sigma,\quad \sigma = 1.36 \pm 0.01.
\end{equation}
Thus, the result confirms that the transition is of first-order from the finite-size scaling (FSS) theory \cite{PRIVMAN-WS-1989,BINDER-RPP-1997}. The exponent $\sigma$ is larger than one, i.e., $\sigma\!>\!1$, however, the anomalous property $C_{X^2}^{\rm max}\to\infty (N\to\infty)$ is evidently seen. 

\begin{figure}[hbt]
\centering
\resizebox{0.49\textwidth}{!}{%
\includegraphics{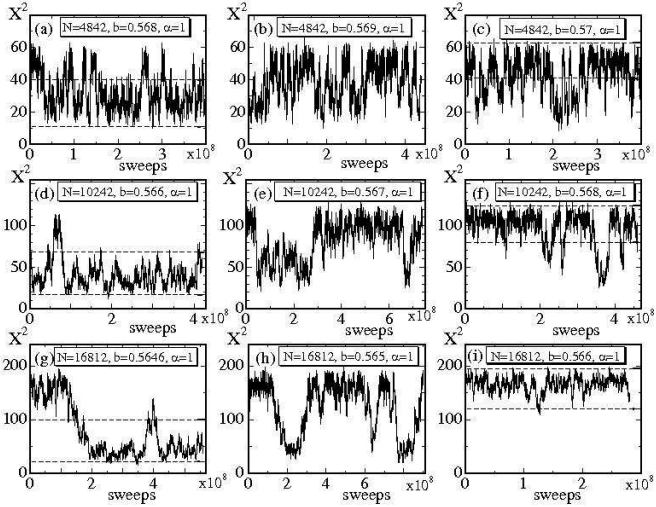}
}
\caption{The variation of $X^2$ versus MCS at the transition region of (a), (b), (c) the $N\!=\!4842$ surface, and (d), (e), (f) the $N\!=\!10242$ surface, and (g), (h), (i) the $N\!=\!16812$ surface. The dashed lines drawn in the figures denote the lower and the upper bounds for the evaluation of the mean values of $X^2$ in the collapsed phase and in the smooth phase.}
\label{fig-3}
\end{figure}
Figures \ref{fig-3}(a)--\ref{fig-3}(i) show the variation of $X^2$ against MCS of model 1 at the transition region of the surfaces $N\!=\!4842$, $N\!=\!10242$, and $N\!=\!16812$. We see in Fig. \ref{fig-3}(e) a jump of $X^2$ from the collapsed phase to the smooth phase and vise versa. If we compute the mean values of $X^2$ in both phases separately, the Hausdorff dimension $H$ can be evaluated by
\begin{equation}
\label{Hausdorff-H}
X^2 \sim N^{2/H}\quad (N\to \infty ).
\end{equation}
The dashed lines drawn horizontally in the figures denote the upper and the lower bounds $X^2_{\rm max}$, $X^2_{\rm min}$ for evaluating the mean values of $X^2$ in each phase such that $X^2_{\rm min} < X^2 < X^2_{\rm max}$. The values of $X^2_{\rm max}$ and  $X^2_{\rm min}$ are shown in Table \ref{table-1}.
\begin{table}[hbt]
\caption{ The lower bound $X^{2 \;{\rm col}}_{\rm min}$ and the upper bound $X^{2 \;{\rm col}}_{\rm max}$ for the mean value $X^2({\rm col})$ in the collapsed state of model 1, and the lower bound $X^{2 \;{\rm smo}}_{\rm min}$ and the upper bound $X^{2 \;{\rm smo}}_{\rm max}$ for the mean value $X^2({\rm smo})$ in the smooth state. }
\label{table-1}
\begin{center}
 \begin{tabular}{cccccc}
  $N$ & $b$& $X^{2 \;{\rm col}}_{\rm min}$ & $X^{2 \;{\rm col}}_{\rm max}$ & $X^{2 \;{\rm smo}}_{\rm min}$ & $X^{2 \;{\rm smo}}_{\rm max}$   \\
 \hline
   4842 & 0.568    & 11 & 40  & --  & -- \\
   4842 & 0.57     & -- & --  & 41  & 63  \\
   10242 & 0.566   & 17 & 68  & --  & -- \\
   10242 & 0.568   & -- & --  & 80  & 124  \\
   16812 & 0.5646  & 22 & 100 & --  & --  \\
   16812 & 0.566   & -- &  -- & 120 & 195  \\
 \hline
 \end{tabular} 
\end{center}
\end{table}

\begin{figure}[hbt]
\centering
\resizebox{0.49\textwidth}{!}{%
\includegraphics{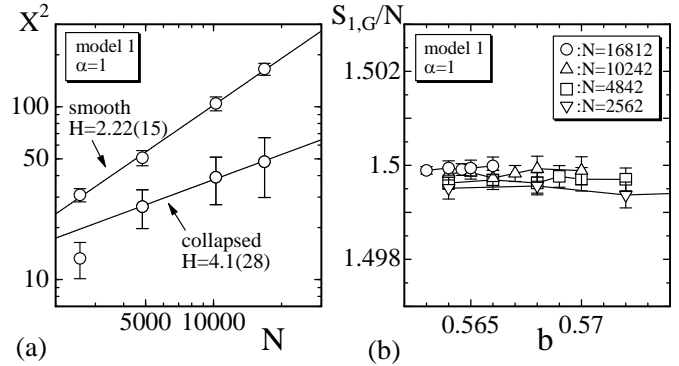}
}
\caption{(a) Log-log plots of $X^2$ vs. $N$ obtained in the smooth phase and in the collapsed phase at the transition point of model 1, (b) the Gaussian bond potential $S_{\rm 1,G}/N$ vs. $b$.}
\label{fig-4}
\end{figure}

The mean values of $X^2$ computed by using the upper and the lower bounds in Table \ref{table-1} are shown in Fig. \ref{fig-4}(a) in a log-log scale. The error bars in Fig. \ref{fig-4}(a) denote the standard deviation obtained from the mean values of $X^2$ and the series $\{X^2\}_i$ with the condition imposed by $X^2_{\rm max}$ and $X^2_{\rm min}$. The straight lines drawn in Fig. \ref{fig-4}(a) are obtained by fitting the data to the scaling relation of Eq. (\ref{Hausdorff-H}). In the collapsed phase, the largest three $X^2$ are used in the fitting. Thus, we have
\begin{eqnarray}
\label{Hausdorff-results-1}
H^{\rm smo}=2.22\pm 0.15, \quad ({\rm smooth} ), \\ H^{\rm col}=4.1\pm 2.8, \quad ({\rm collapsed} ), \qquad ({\rm model \;1}).\nonumber
\end{eqnarray}
In the smooth phase, we have an expected result, which is close to the topological dimension $H\!=\!2$, while we have an unphysical $H (>3)$ in the collapsed phase. In the case of $\alpha\!=\!0$ in \cite{KOIB-PLA2007}, we have a physical $H$ even in the collapsed phase. Therefore, the result $H^{\rm col} (>3)$ in Eq. (\ref{Hausdorff-results-1}) allows us to understand that the in-pane share elasticity strengthens the transition at $\alpha\!=\!0$. Our model is not self-avoiding, consequently, the size of the surface in the collapsed phase shrinks if the transition is strong. As a consequence, the Hausdorff dimension becomes unphysical in the collapsed phase. 
 
Figure  \ref{fig-4}(b) shows the Gaussian bond potential $S_{\rm 1,G}/N$ vs. $b$. From the scale invariant property of the partition function, the relation $S_{\rm 1,G}/N\!=\!3/2$ should hold in the whole region of $b$ and $\alpha$. In our simulations, the expected relation is satisfied as we see in Fig. \ref{fig-4}(b). Therefore, we consider that the simulations were correctly performed.

Here we should note on the reason why $\sigma$ in Eq. (\ref{scaling-CX2}) is larger than $\sigma\!=\!1$. The reason seems to be the low statistics of the simulations on the large sized surfaces. We see from Fig. \ref{fig-3}(e) that $n_{\rm tot}$ should be $n_{\rm tot}>4\times10^8$ or more even on the $N\!=\!10242$ surface. This implies that the high statistics simulations for surface models are not so easy even on the surface of size $N\!=\!10242$. Nevertheless, we must emphasize that the first-order nature of the transition is conclusive from the numerical results in this paper.  

\begin{figure}[hbt]
\centering
\resizebox{0.49\textwidth}{!}{%
\includegraphics{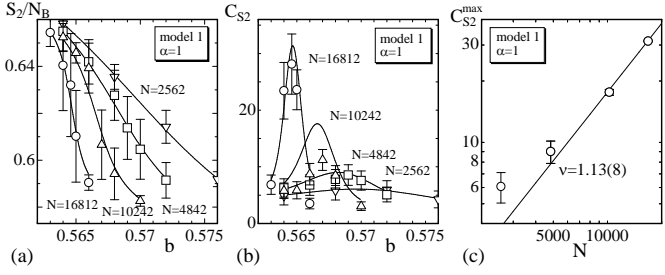}
}
\caption{(a) The one-dimensional bending energy $S_2/N_B$ versus $b$ of model 1, (b) the specific heat $C_{S_2}$ versus $b$, and (c) log-log plots of the peak $C_{S_2}^{\rm max}$ against $N$. The straight line in (c) is obtained by fitting the largest three data to Eq. (\ref{scaling-CS2}). The critical exponent $\nu\!=\!1.13(8)$ indicates that the transition of surface fluctuations is of first-order.}
\label{fig-5}
\end{figure}
The bending energy $S_2/N_B$, which is defined in Eq. (\ref{Disc-Eneg}), is plotted in Fig. \ref{fig-5}(a), where $N_B$ is the total number of bonds. We also see a rapid change in $S_2/N_B$ just like in $X^2$ in Fig. \ref{fig-2}(a). This suggests a discontinuous transition of surface fluctuations. 

The specific heat $C_{S_2}$ for $S_2$ is defined by 
\begin{equation}
\label{specific-heat}
C_{S_2} = {b^2\over N} \langle \; \left( S_2 \!-\! \langle S_2 \rangle\right)^2\rangle,
\end{equation}
and is plotted in Fig. \ref{fig-5}(b) against $b$. An anomalous peak $C_{S_2}^{\rm max}$ can also be expected in $C_{S_2}$ just as in $C_{X^2}$, and the peaks are apparently seen in Fig. \ref{fig-5}(b). Figure \ref{fig-5}(c) shows a log-log plot of $C_{S_2}^{\rm max}$ against $N$. The straight line is obtained by fitting the largest three data, and we have a critical exponent such that
\begin{equation}
\label{scaling-CS2}
C_{S_2}^{\rm max} \propto N^\nu, \quad \nu=1.13\pm 0.08,\quad ({\rm model\; 1}).
\end{equation}
The result indicates that the transition of surface fluctuations is of first-order, because $\nu$ almost satisfies $\nu\!=\!1$. This is a remarkable result distinguishing model 1 in this paper from the model in \cite{KOIB-PLA2007}, where the transition of surface fluctuations is reported to be of second-order. Therefore, we understand that the in-plane shear elasticity strengthens the transition of surface fluctuations in the surface model with one-dimensional bending energy.  

\begin{figure}[hbt]
\centering
\resizebox{0.49\textwidth}{!}{%
\includegraphics{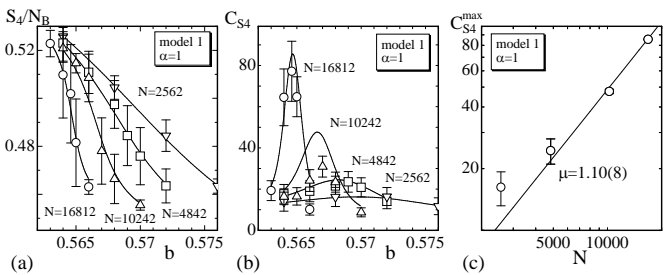}
}
\caption{(a) The two-dimensional bending energy $S_4/N_B$ versus $b$ of model 1, (b) the variance $C_{S_4}$ of $S_4$ versus $b$, and (c) log-log plots of the peak $C_{S_4}^{\rm max}$ against $N$. The straight line in (c) is obtained by fitting the largest three data to  $C_{S_4}^{\rm max}\sim N^\mu$, and the result $\nu\!=\!1.10(8)$ indicates that the transition of surface fluctuations is of first-order.}
\label{fig-6}
\end{figure}
In order to confirm more convincingly this fact that the model undergoes a first-order transition of surface fluctuations, we plot in Fig. \ref{fig-6}(a) the two-dimensional bending energy defined by 
\begin{eqnarray}
\label{two-dim-bending} 
 S_4=\sum_{(ij)}\left(1-{\bf n}_i\cdot {\bf n}_j\right),
\end{eqnarray} 
where ${\bf n}_i$ is a unit normal vector of the triangle $i$. Although the bending energy $S_4$ is not included in the Hamiltonian, it represents how large the surface fluctuates. The variance $C_{S_4}\!=\!(1/ N) \langle \; \left( S_4 \!-\! \langle S_4 \rangle\right)^2\rangle$ of $S_4$ is plotted in Fig. \ref{fig-6}(b), and we see that $C_{S_4}$ has an expected anomalous peak, which reflects a discontinuous nature of $S_4$. Figure \ref{fig-6}(c) is a log-log plot of the peak value $C_{S_4}^{\rm max}$ against $N$, where the straight line is drawn by the fitting the data according to the scaling relation $C_{S_4}^{\rm max}\sim N^\mu$, and we have a critical exponent  $\mu = 1.10\pm0.08$. Then, we again find that the transition of surface fluctuations is of first-order from the obtained exponent and the FSS theory.

\begin{figure}[htb]
\centering
\resizebox{0.49\textwidth}{!}{%
\includegraphics{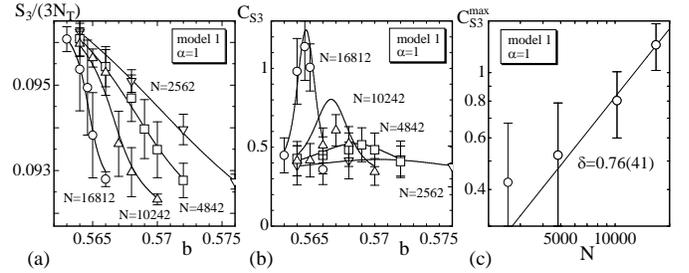}
}
\caption{(a) The in-plane shear energy $S_3/(3N_T)$ versus $b$, (b) the variance $C_{S_3}$ of $S_3$ versus $b$, and (c) log-log plots of the peak $C_{S_3}^{\rm max}$ against $N$. The straight line in (c) is obtained by fitting the largest three data to  $C_{S_3}^{\rm max}\sim N^\delta$.}
\label{fig-7}
\end{figure}
Finally in this subsection, we show the in-plane energy $S_3/(3N_T)$ versus $b$, the variance $C_{S_3}$ of $S_3$ versus $b$, and the log-log plot of the peak value $C_{S_3}^{\rm max}$ against $N$, in Fig. \ref{fig-7}(a), Fig. \ref{fig-7}(b), and Fig. \ref{fig-7}(c), respectively. An anomalous peak can also be seen in $C_{S_3}$. The scaling behavior can be seen in Fig. \ref{fig-7}(c), and we have 
\begin{equation}
\label{scaling-CS3}
C_{S_3}^{\rm max}\sim N^\delta,\quad \delta=0.76\pm0.41.
\end{equation}
This implies that the {\it in-plane order-disorder} transition is of first-order because $\delta$ can be seen as $\delta\!=\! 1$ within the error, however this conclusion is less accurate because of the large errors. 

The in-plane order-disorder transition is expected to be discontinuous at least when $\alpha\!=\!1$ in Eq. (\ref{Part-Func}), and both of the collapsing transition and the transition of surface fluctuations are of first-order. Moreover, we expect that the order of the in-plane order-disorder transition changes according to the value $\alpha$. The transition is expected to change from the discontinuous one to a continuous one at certain value of $\alpha$ at least in $0\!<\!\alpha\!<\!1$.

\subsection{Nambu-Goto potential model}
The numerical data obtained from model 2 are presented in this subsection. The presentation is almost parallel to the one of model 1 in the previous subsection. The parameter $\alpha$ in Eq. (\ref{Part-Func}) is fixed to $\alpha\!=\!1$ just as in model 1.

\begin{figure}[hbt]
\centering
\resizebox{0.49\textwidth}{!}{%
\includegraphics{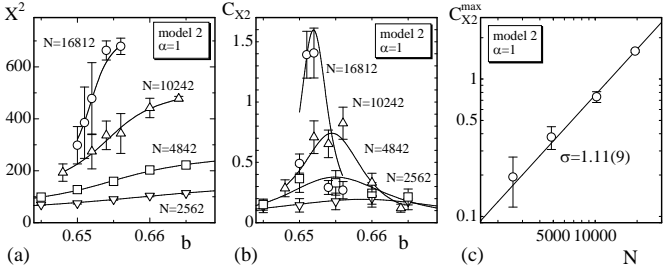}
}
\caption{(a) The mean square size $X^2$ versus $b$ of model 2, (b) the variance $C_{X^2}$ of $X^2$ versus $b$, and (c) log-log plots of the peak $C_{X^2}^{\rm max}$ against $N$. The solid lines in (a) and (b) are drawn by the multihistogram reweighting technique \cite{Janke-histogram-2002}. The straight line in (c) is obtained by fitting the data to Eq. (\ref{scaling-CX2}). The critical exponent $\sigma\!=\!1.11(9)$ indicates that the collapsing transition is of first-order.}
\label{fig-8}
\end{figure}
Figure \ref{fig-8}(a) shows the mean square size $X^2$ defined in Eq. (\ref{X2}). The solid curves are also drawn by the multihistogram technique like those in the figures in the previous section. The variance $C_{X2}$ is shown in Fig. \ref{fig-8}(b), and the peak values  $C_{X2}^{\rm max}$ are plotted in Fig. \ref{fig-8}(c) against $N$ in a log-log scale. The straight line is drawn by fitting all of the four data to Eq. (\ref{scaling-CX2}). We have the critical exponent $\sigma\!=\!1.11\!\pm\! 0.09$, which implies a first-order collapsing transition. 

We should remark that the in-plane energy $S_3$ decreases the multitude of phases of the model in \cite{KOIB-EPJB2007-3}, and only two phases; the smooth and the collapsed phases, survive. These two phases always seen in the standard surface models such as those of \cite{KOIB-PRE-20045-NPB-2006} and in the mesh work model of \cite{KOIB-PLA2007}. The reason for the shrinkage in the multitude of phases in the model of \cite{KOIB-EPJB2007-3} is intuitively understood; in fact, the linear and the planar surfaces are constructed not only from regular triangles but also from oblong triangles. To the contrary, oblong triangles are prohibited to occur in model 2 of this paper due to the presence of $S_3$. 

\begin{figure}[hbt]
\centering
\resizebox{0.49\textwidth}{!}{%
\includegraphics{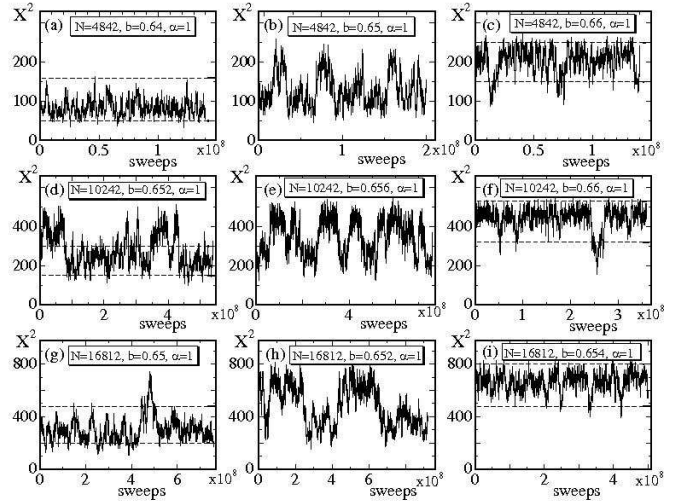}
}
\caption{The variation of $X^2$ versus MCS of model 2 at the transition region of (a), (b), (c) the $N\!=\!4842$ surface, and (d), (e), (f) the $N\!=\!10242$ surface, and (g), (h), (i) the $N\!=\!16812$ surface. The dashed lines drawn on the figures denote the lower and the upper bounds for the mean values of $X^2$ in the collapsed phase and in the smooth phase.}
\label{fig-9}
\end{figure}
Figures \ref{fig-9}(a)--\ref{fig-9}(i) show the variation of $X^2$ against MCS of model 2, where $N\!=\!4842$, $N\!=\!10242$, and $N\!=\!16812$. It is clear that two different states coexist; one is characterized by the large $X^2$ and the other by the small $X^2$, which respectively correspond to the smooth phase and the collapsed state. The transition, which separates one state from the other, can be called the collapsing transition just as in model 1.

\begin{figure}[hbt]
\centering
\resizebox{0.49\textwidth}{!}{%
\includegraphics{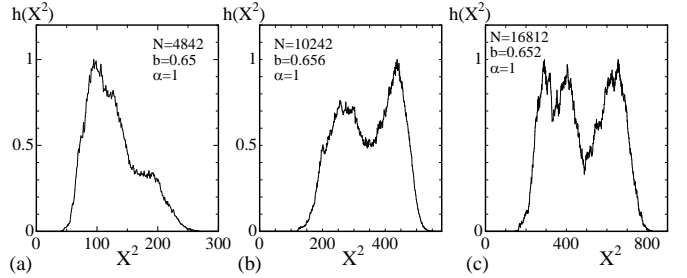}
}
\caption{The distribution (or histogram) $h(X^2)$ of $X^2$ at the transition point of the surface of size (a) $N\!=\!4842$, (b) $N\!=\!10242$, and (c) $N\!=\!16812$ of model 2.}
\label{fig-10}
\end{figure}
In order to see the first-order nature of the collapsing transition more clearly, we show in Figs. \ref{fig-10}(a)--\ref{fig-10}(c) the normalized distribution (or histogram) $h(X^2)$ of $X^2$ at the transition point of the surfaces of $N\!=\!4842$,  $N\!=\!10242$, and $N\!=\!16812$. The histograms in Figs. \ref{fig-10}(a)--\ref{fig-10}(c) correspond to the variations of $X^2$ in Figs. \ref{fig-9}(e), and \ref{fig-9}(h), respectively. The double peak structure can be seen in $h(X^2)$ of the $N\!=\!10242$ surface and the $N\!=\!16812$ surface, and this clearly indicates that the collapsing transition is of first-order.

It is interesting to see whether the collapsing transition is physical or not, i.e., whether or not the Hausdorff dimension $H$ is less than the physical bound in the collapsed phase. The model is allowed to self-intersect like model 1 in the previous section, however, the transition is considered to be physical if both of $H$ in the smooth phase and the collapsed phase are less than the physical bound; $H\!<\!3$. In Table \ref{table-2}, we show the lower and the upper bounds for the mean value $X^2$, from which we have $H$ by using Eq. (\ref{Hausdorff-H}). The values shown in Table \ref{table-2} are indicated by the horizontal dashed lines in Figs. \ref{fig-9}(a)--\ref{fig-9}(i).
\begin{table}[hbt]
\caption{ The lower bound $X^{2 \;{\rm col}}_{\rm min}$ and the upper bound $X^{2 \;{\rm col}}_{\rm max}$ for the mean value $X^2({\rm col})$ in the collapsed state of model 2, and the lower bound $X^{2 \;{\rm smo}}_{\rm min}$ and the upper bound $X^{2 \;{\rm smo}}_{\rm max}$ for the mean value $X^2({\rm smo})$ in the smooth state. These values correspond to the horizontal dashed lines in Figs. \ref{fig-9}(a)--\ref{fig-9}(i). }
\label{table-2}
\begin{center}
 \begin{tabular}{cccccc}
  $N$ & $b$& $X^{2 \;{\rm col}}_{\rm min}$ & $X^{2 \;{\rm col}}_{\rm max}$ & $X^{2 \;{\rm smo}}_{\rm min}$ & $X^{2 \;{\rm smo}}_{\rm max}$   \\
 \hline
   4842 & 0.64    & 50  & 160 & --   & -- \\
   4842 & 0.66    & --  & --  & 150  & 250  \\
   10242 & 0.652  & 150 & 300 & --   & --  \\
   10242 & 0.66   & --  &  -- & 320  & 530  \\
   16812 & 0.65   & 200 & 480 & --   & --  \\
   16812 & 0.654  & --  &  -- & 480  & 800  \\
 \hline
 \end{tabular} 
\end{center}
\end{table}

\begin{figure}[hbt]
\centering
\resizebox{0.49\textwidth}{!}{%
\includegraphics{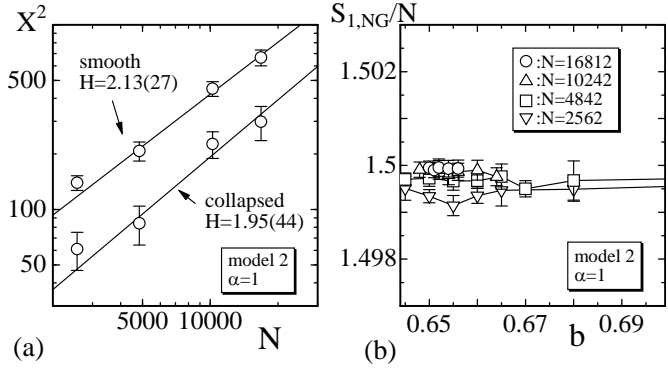}
}
\caption{(a) Log-log plots of $X^2$ vs. $N$ obtained in the smooth phase and in the collapsed phase at the transition point, (b) the Nambu-Goto potential $S_{\rm 1,NG}/N$ vs. $b$.  The fittings in (a) were done by using the largest three data both in the collapsed phase and in the smooth phase. }
\label{fig-11}
\end{figure}
Figure \ref{fig-11}(a) shows log-log plots of the mean values $X^2$ against $N$, where $X^2$ were obtained from the variation of $X^2$ shown in  Figs. \ref{fig-9}(a)--\ref{fig-9}(i) by using $X^2_{\rm max}$ and $X^2_{\rm min}$ in Table \ref{table-2}. The straight lines were drawn by fitting the largest three data to Eq. (\ref{Hausdorff-H}) both in the smooth phase and in the collapsed phase. We have 
\begin{eqnarray}
\label{Hausdorff-results-2}
&&H^{\rm smo}=2.13\pm 0.27, \quad ({\rm smooth} ), \\ &&H^{\rm col}=1.95\pm0.44, \quad ({\rm collapsed} ), \qquad ({\rm model\; 2}). \nonumber
\end{eqnarray}
Thus, the collapsing transition of model 2 is considered to be physical, because $H^{\rm col}$ is less than the physical bound. This point distinguishes model 2 from model 1, whose collapsing transition is considered as unphysical at least when $\alpha\!=\!1$ as demonstrated in the previous subsection. 

The Nambu-Goto potential $S_{\rm 1,NG}/N$  is plotted in Fig. \ref{fig-11}(b). We find that the expected relation $S_{\rm 1,NG}/N\!=\!3/2$ is satisfied. This implies that model 2 is well-defined and moreover that the simulations of model 2 are successfully performed.

\begin{figure}[hbt]
\centering
\resizebox{0.49\textwidth}{!}{%
\includegraphics{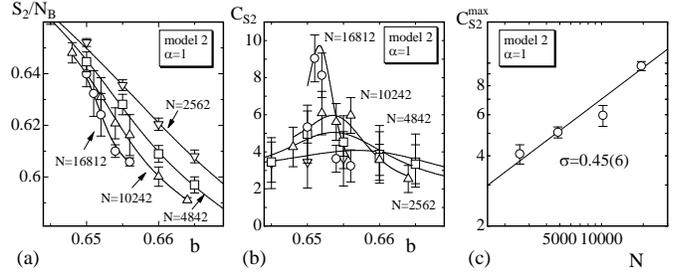}
}
\caption{(a) The one-dimensional bending energy $S_2/N_B$ vs. $b$ of model 2, (b) the specific heat $C_{S_2}$ versus $b$, and (c) log-log plots of the peak $C_{S_2}^{\rm max}$ against $N$. The straight line in (c) is obtained by fitting the data to Eq. (\ref{scaling-CS2}). The critical exponent $\nu\!=\!0.45(6)$ indicates that the transition of surface fluctuations is of second-order.}
\label{fig-12}
\end{figure}
Finally in this subsection, we plot $S_2/N_B$ vs. $b$ in Fig. \ref{fig-12}(a) and the specific heat $C_{S_2}$ vs. $b$ in Fig. \ref{fig-12}(b). The peaks $C_{S_2}^{\rm max}$  of the specific heat seen in  Fig. \ref{fig-12}(b) imply that model 2 undergoes a phase transition of surface fluctuations. The log-log plot of $C_{S_2}^{\rm max}$ vs. $N$ is shown in Fig. \ref{fig-12}(c). The exponent $\nu$ defined by Eq. (\ref{scaling-CS2}) is obtained as follows:
\begin{equation}
\label{nu-model-2}
\nu=0.45\pm 0.06,\qquad ({\rm model \; 2}).
\end{equation}
This value of $\nu$ implies that the transition of surface fluctuations is of second-order at $\alpha\!=\!1$. However, a possibility of the first-order transition is not completely eliminated. Large scale simulations are still necessary to clarify this point. 

We performed the simulations for model 2 with $\alpha\!=\!0.5$ by using the same sized surfaces as those for the case $\alpha\!=\!1$ presented above. The phase structure is almost identical to the case $\alpha\!=\!1$ except the strength of the transitions. The collapsing transition is considered to be weakened but it still remains discontinuous at $\alpha\!=\!0.5$, because the finite size-scaling analysis of the peak values of $C_{X^2}^{\rm max}$ indicates a first-order collapsing transition. However, the transition of surface fluctuations changes to a higher-order one and almost disappears. Thus, we find that the strength of the transitions changes depending on the value of $\alpha$ in model 2 as well as in model 1.

\section{Summary and conclusions}
In this article, we have investigated how the in-plane shear elasticity influences the transitions observed in two-types of the meshwork models in \cite{KOIB-PLA2007} and \cite{KOIB-EPJB2007-3}. The models are described as follows: The surface shape of the first model in \cite{KOIB-PLA2007} is maintained by a one-dimensional bending energy and by the Gaussian bond potential, while the surface shape of the second model in \cite{KOIB-EPJB2007-3} is maintained by the one-dimensional bending energy and by the Nambu-Goto potential. One of the transitions in the model of \cite{KOIB-PLA2007} is called the collapsing transition, which is of first-order and separates the smooth phase from the collapsed phase at finite bending rigidity. The other is a continuous one called the transition of surface fluctuations. We should remark that these two transitions are of discontinuous in the conventional surface model with two-dimensional bending energy \cite{KOIB-PRE-20045-NPB-2006}. The second model studied in \cite{KOIB-EPJB2007-3} has a variety of phases including the smooth, the planar, and the linear phases in contrast to the first model in \cite{KOIB-PLA2007}. 

The reason why we study in this paper the meshwork models of \cite{KOIB-PLA2007} and \cite{KOIB-EPJB2007-3} is that the models have no in-plane resistance force against the in-plane deformations. The models have a resistance only against the bending deformation and the tensile deformation. The linear surface of the model in  \cite{KOIB-EPJB2007-3} consists of oblong surfaces, and the planar surface consists of both regular triangles and oblong ones. No energy cost is necessary for the in-plane deformations in both of the models in \cite{KOIB-PLA2007,KOIB-EPJB2007-3}.  For this reason, we expect that the in-plane energy makes a non-trivial effect on the transitions. We call the first and the second models with the in-plane energy as model 1 and model 2, respectively. The value of $\alpha$ the coefficient of the in-plane energy was assumed as  $\alpha\!=\!1$ in the simulations in both models.

Our numerical results of model 1 show that the transition of surface fluctuations is of first-order, which was reported to be of second-order in the model without the in-plane shear energy just stated as above. The collapsing transition is also strengthened, although the order of the transition is identical to the case without the in-plane energy. Therefore, we conclude that the in-plane shear energy can strengthen the transitions observed in the meshwork model with the Gaussian bond potential.

Moreover, we find in model 1 that the in-plane order-disorder transition is of first-order; the in-plane energy discontinuously changes against the bending rigidity at the transition point, where both of the one-dimensional and the two-dimensional bending energies also discontinuously change. The first-order nature of the in-plane order-disorder transition is not so accurate because of the large errors in the critical exponent for the variance of the in-plane energy. 

In the case of model 2, the variety of phases seen in the model in \cite{KOIB-EPJB2007-3} disappears, and the phase structure is almost identical to that of model 1. The reason of this is because oblong triangles are suppressed due to the presence of the in-plane energy. The collapsing transition between the smooth phase and the collapsed phase is of first-order. Moreover, this transition is considered as physical, because the Hausdorff dimension $H$ is less than the physical bound; $H\!<\!3$, even in the collapsed phase close to the transition point. 
The transition of surface fluctuations of model 2 is considered to be continuous. The finite-size scaling analysis of the specific heat of the one-dimensional bending energy supports this conclusion.  

We should note that the strength of the transitions changes depending on the value of $\alpha$. One can also expect that the unphysical collapsing transition of model 1 changes to the physical one at some values of $\alpha$ in the range $0\!<\!\alpha\!<\!1$, because the strength of the transitions weakens with decreasing $\alpha$. We know that the collapsing transition of model 1 is physical in the limit of $\alpha\to 0$ \cite{KOIB-PLA2007}.
It is also possible that the transition of surface fluctuations in model 2 turns to be a discontinuous one with increasing $\alpha$. It is interesting to study the conventional curvature surface model in \cite{KOIB-PRE-20045-NPB-2006} by including the in-plane shear energy in the Hamiltonian.

This work is supported in part by a Grant-in-Aid for Scientific Research from Japan Society for the Promotion of Science. The author I.E. acknowledges Kaneyama Ltd. for a financial support.




\end{document}